\documentclass[preprint]{aastex63}

\newcommand\Alfven{Alfv$\acute{e}$n}
\newcommand\Rsun{R$_\odot$}

\received{June 3, 2019}
\accepted{April 7, 2020}
\usepackage{CJK}

\submitjournal{Frontiers}

\shorttitle{Two Type II Radio Bursts}
\shortauthors{Ma \& Chen}

\begin{document}

\title{Two Successive Type II Radio Bursts Associated with B-class Flares and Slow CMEs}

\correspondingauthor{Suli Ma}
\email{sma@nao.cas.cn}

\begin{CJK*}{UTF8}{gbsn}
\author[0000-0002-5431-6065]{Suli Ma (马素丽)}
\affiliation{Key Laboratory of Solar Activity, National Astronomical Observatories,\\
Chinese Academy of Sciences, Beijing, 100012, China}

\author[0000-0001-6076-9370]{Huadong Chen (陈华东)}
\affiliation{Key Laboratory of Solar Activity, National Astronomical Observatories,\\
Chinese Academy of Sciences, Beijing, 100012, China}
\affiliation{School of Astronomy and Space Science, \\
University of Chinese Academy of Sciences, Beijing, 100049, China}
\begin{abstract}

From 2018 Oct 12 to 13, three successive solar eruptions (E1--E3) with B-class flares and poor white light coronal mass ejections (CMEs) occurred from the same active region NOAA AR 12724. Interestingly, the first two eruptions are associated with Type II radio bursts but the third is not. Using the soft X-ray flux data, radio dynamic spectra and dual perspective EUV intensity images, we comparatively investigate the three events. Our results show that their relevant flares are weak (B2.1, B7.9 and B2.3) and short-lived (13, 9 and 14 minutes). The main eruption directions of E1 and E2 are along $\sim$45$^\circ$ north of their radial directions, while E3 primarily propagated along the radial direction. In the EUV channels, the early speeds of the first two CMEs have apparent speeds of $\sim$320 km s$^{-1}$ and $\sim$380 km s$^{-1}$, which could exceed their respective local \Alfven\ speeds of $\sim$300 km s$^{-1}$ and $\sim$350 km s$^{-1}$. However, the CME in the third eruption possesses a much lower speed of $\sim$160 km s$^{-1}$. These results suggest that the observed Type II radio bursts in the eruptions E1 and E2 are likely triggered by their associated CMEs and the direction of eruption and the ambient plasma and magnetic environments may take an important place in producing Type II radio burst or shock as well.

\end{abstract}



\section{Introduction} 
Solar Type II radio bursts were first reported by \cite{PayneScott47} and named by \citet{Wild50}. Usually, a Type II burst appeared as a slowly drifting, from high to low frequencies, narrow frequency band patterns in radio dynamic spectra \citep[e.g.,][]{Nelson85}. It is generally believed that type II bursts are excited by magnetohydrodynamics (MHD) shocks in the solar atmosphere \citep[e.g.,][and references therein]{Nelson85,Cliver99,Vrsnak08,Nindos08,Nindos11}.  
 
Since the first discovery of Type II bursts, they have been found to be closely related to both flares \citep[e.g.,][]{Wild54,Maxwell62,Dodge75} and high-velocity ejections \citep[e.g.,][]{Dodson53,Giovanelli58,Swarup60}. 
The blast wave initiated by flares and piston-driven mechanism associated with coronal mass ejections (CMEs) have become the two main competitors of the triggering mechanism of the Type II bursts as reviewed by the papers \citep[e.g.,][]{Cliver99,Vrsnak08,Nindos11} and also suggested by some recent studies \citep[e.g.,][]{Zheng18,Frassati19,Eselevich19}. 
In the piston-driven mechanism, besides CME front, some other triggers, such as soft X-ray jet, erupting coronal loop and eruptive magnetic flux rope are also proposed to explain the production of shock or Type II burst \citep[e.g.,][]{Klein99,Klassen03,Dauphin06,Su15,Eselevich17,Grechnev18}. 

In theory, the occurrence of a coronal shock requires the introduction of a sudden disturbance in the corona, which should travel with a speed faster than the local \Alfven\ velocity. Early studies \citep[e.g.,][]{Gosling76} indicate that the velocities of CMEs with type II bursts would exceed 400-550 km s$^{-1}$. 
However, \citet{Gopalswamy01} made a statistical study and found that 50\% of limb CMEs associated Type II bursts during 1995 to 1997 have speeds lower than 500 km s$^{-1}$ and the lower cutoff of these CMEs' speeds may reach $\sim$ 250 km s$^{-1}$. So far, detailed case studies about Type II radio bursts with slow CMEs and weak flares (below C-class) have been very rare. 

In this paper, we present a case study about two successive Type II bursts associated with B-class flares and CMEs with slow speeds below $\sim$400 km s$^{-1}$ in the period of solar activity minimum. 
Our comparative investigations suggest that these Type II radio bursts are likely triggered by their associated CMEs and we also discuss the influence from the ambient coronal magnetic structures on the eruptions.

\section{Observations}
From 2018 Oct 12 to 13, three eruptions orderly took place in the active region (AR) NOAA 12724.
AR 12724 is near the solar east limb in the field of view (FOV) of the ground-based telescopes (spectrographs) or space-based telescopes on geosynchronous satellites, such as the Solar and Heliospheric Observatory $SOHO$ and the Solar Dynamics Observatory \citep[$SDO$,][]{Pesnell12}. 
The Solar Terrestrial Relations Observatory \citep[STEREO,][]{Kaiser08} consists of two space-based observatories -- one ahead of Earth in its orbit ($STA$), the other trailing behind ($STB$, lost communications since 2014 Oct 1). The $STA$ orbits the Sun with a radius slightly smaller than 1AU and the separation angle between $STA$ and the Earth was about 105$^\circ$ during the three events. The host AR is located on the disk in the view of $STA$. 

We use the dynamic spectrum data from the radio spectrograph Observation Radio Frequence pour l’Etude des Eruptions Solaires (ORFEES) observing between 140 and 1000 MHz, Learmonth solar radio spectrograph covering a frequency range of 25--180 MHz \citep[LEAR,][]{Kennewell03}, and the CALLISTO spectrometer \citep{Benz09} at the Greenland Observatory.  
Intensity images provided by the Atmospheric Imaging Assembly \citep[AIA,][]{Lemen12} on SDO and the EUV Imager \citep[EUVI][]{Wuelser04} in the Sun Earth Connection Coronal and Heliospheric Investigation \citep[$SECHHI$;][]{Howard08} on $STA$ are also utilized to study the early stages of the eruptions. 
The observation from the Large Angle and Spectrometric Coronagraph \citep[LASCO;][]{Brueckner95} on-board SOHO with a FOV of 2--6.0 R$_\odot$ help us to check the associated CMEs.

\section{Results}
\subsection{Overview of the Events}
The general information of the three successive eruption events (E1--E3) are listed in Table~1. 
All the three events originated from the same active region (AR) NOAA~12724 and each of them involved a filament eruption, a B-class flare and a slow faint white light (WL) CME. 
Type II radio bursts only appeared in E1 and E2.
The detailed magnitude and the start, peak, and end time of each flare are given in the ``Flare Class'' and ``Flare Time'' column, respectively.
The central locations of the relevant filaments (F1--F3) in the FOV of AIA are presented in the ``Filament Center'' column.
The CMEs' angular widths (AWs) and their quadratic speeds (obtained by performing second-order polynomial fittings to the height-time measurements) at the final height measurements are placed in the ``CME AW'' and ``CME Speed'' columns, respectively.
 \begin{table}[htp]
\caption{General Information of the Eruptions} 
\begin{tabular}{ccccccccc}
 \hline
 \hline
Event &  Date &  Flare &  Flare Time (UT) & Radio & Filament& CME AW  & CME Speed  \\
 &&Class& start  peak  end &Burst& Center & (deg) & (km/s)\\
\hline
E1 & 2018 Oct 12 & B2.1 & 01:43   01:50   01:56 & II     & -900\arcsec,-150\arcsec&  46 &333 \\ 
E2 & 2018 Oct 12 & B7.1 & 14:04   14:08   14:13 & II,IIIs,IV &-850\arcsec,-160\arcsec & 44 &492 \\ 
E3 & 2018 Oct 13 & B2.3 & 13:28   13:34   13:42 & III  & -800\arcsec,-240\arcsec & 38 & 133 \\ 
\hline
\end{tabular}
\end{table}
\subsection{Filaments}

The AIA 304 \AA\ (left column) and EUVI A 195 \AA\ (right column) images in Figure~1 show the morphologies of the AR 12724, the erupting filaments (F1--F3) and their corresponding flares (Flare1--Flare3). 
In the 304 \AA\ images, it can be seen that F1 and F2 almost have a north-south orientation, while F3 runs from east to west.
According to the EUVI A 195 \AA\ observations, Flare1 and Flare2 were located in the northwest of AR 12724, while Flare3 mainly lay in the AR's southeast. To the east of AR 12724, a small emerging active region labelled ``New'' appeared to have nothing to do with the eruptions.\\

\begin{figure}[ht!]
\begin{center}
\includegraphics[width=15cm]{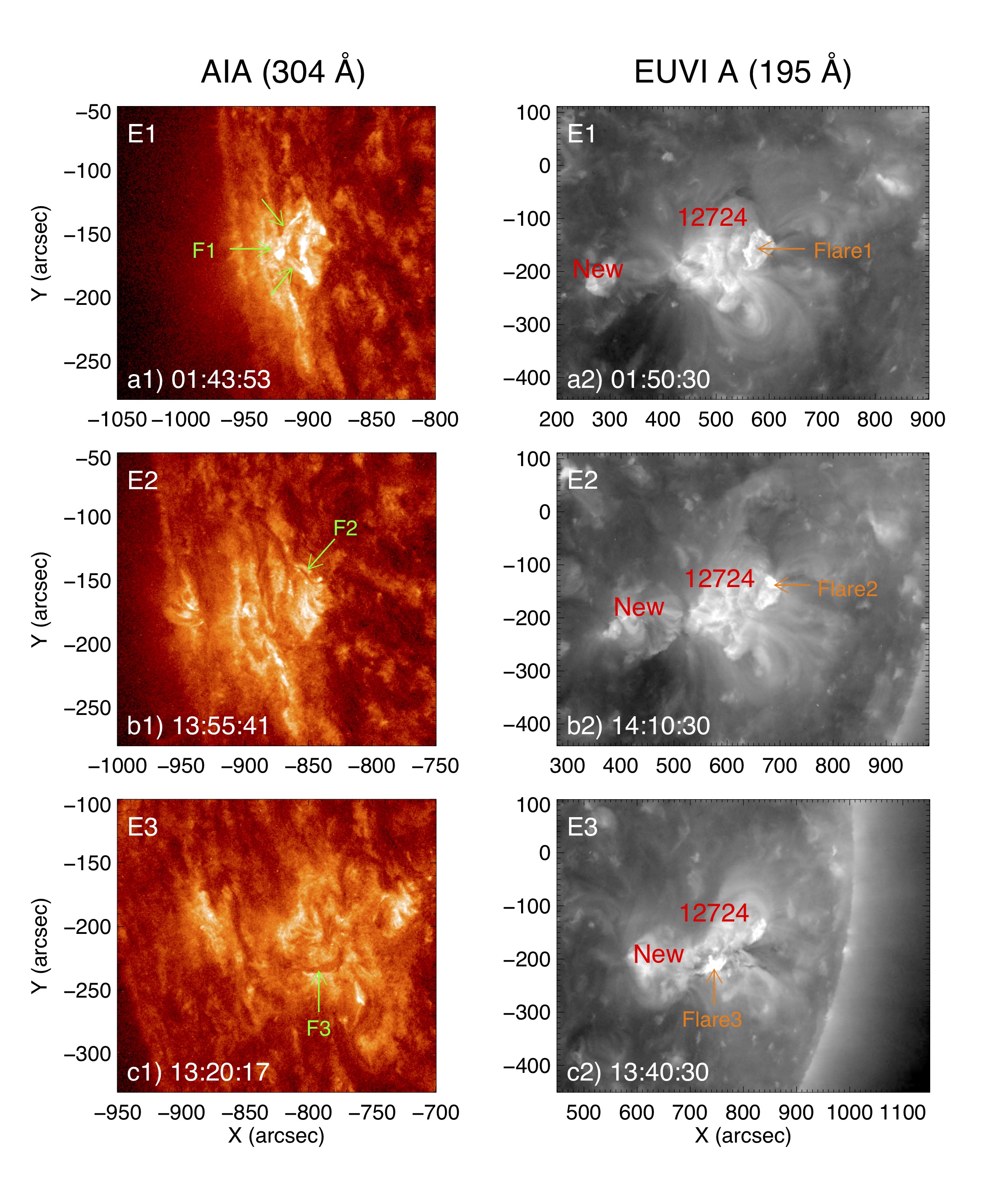}
\end{center}
\caption{Observations in EUV passbands displaying the morphology of the host active region NOAA 12724 during the occurrences of E1(a1-a2), E2(b1,b2) and E3(c1,c2). The images in left column show the AIA observation at 304 \AA\ and the green arrows point to the locations of filament1 (F1), filament2 (F2) and filament3 (F3). The images in right column are EUVI images at 195 \AA\ and the orange arrows direct the positions of Flare1, Flare2 and Flare3.} \label{fig:1}
\end{figure}

\subsection{Flares and Radio Bursts}
The GOES X-ray fluxes and radio dynamic spectrums associated with E1, E2 and E3 are shown in the panels (a) and (b) of Figures~2-4, respectively. Figure~2(a) displays the short-lived B2.1 Flare1.
In Figure~2(b), a type II radio burst can be found in the Learmonth's radio dynamic spectrum, suggesting that a shock (Shock1) was generated during E1. The type II radio burst, including a fundamental band (F) and a harmonic band (H) with two splitting lanes (H$_{L}$ and H$_{U}$), started from $\sim$01:52 UT, when Flare1 has entered its descending phase.
After $\sim$02:00 UT, it gradually disappeared.
The frequencies along F, H$_{L}$ and H$_{U}$ change from about 40, 78, and 80 MHz to 26, 54, and 64 MHz, respectively. 
Their average frequency drift rates are -0.086 MHz s$^{-1}$, -0.087 MHz s$^{-1}$ and -0.109 MHz s$^{-1}$. 
We adopt the frequency values represented by the black (F), red (H$_{L}$) and blue (H$_{U}$) dashed curves to measure local plasma densities and further derive the speeds of Shock1 (see Section~3.6).


It is generally believed that the band-splitting is caused by the emission from the upstream and downstream shock regions and the downstream/upstream density jump ($X$) could provide an estimate of the coronal \Alfven\ speed \citep[e.g,][]{Smerd74,Mann95,Vrsnak01}.
The density jump can be described as 
    \begin{equation}
    X = \frac {n_2}{n_1}= (\frac{f{_U}}{f{_L}})^2
    \end{equation} 
    \citep{Vrsnak02}.
Here, $n_1$ and $n_2$ are the electron densities of the plasma at the frequency $f_L$ in the lower frequency branch and at the frequency $f_U$ in the upper frequency branch of the harmonic bands, respectively. 
For the Type II radio burst in E1, we take $f{_U} = 77.5$ and $f{_L} = 64.5$ at 01:54:32 UT (indicated by the vertical line in Figure~2(b)) and obtain $X = 1.44$, indicating that Shock1 is a weak shock. 
Under the quasi-perpendicular shock approximation and a plasma beta $\beta->0$, the Alfv$\acute{e}$n Mach number $M_{A}$ is related to the compression $X$ as 
\begin{equation}
M_{A}=\sqrt[]{\frac{X(X+5)}{2(4-X)}}
\end{equation}
Then, $M_{A}$ around 1.35 can be derived for Shock1.
 
\begin{figure}[ht!]
\begin{center}
\includegraphics[width=15cm]{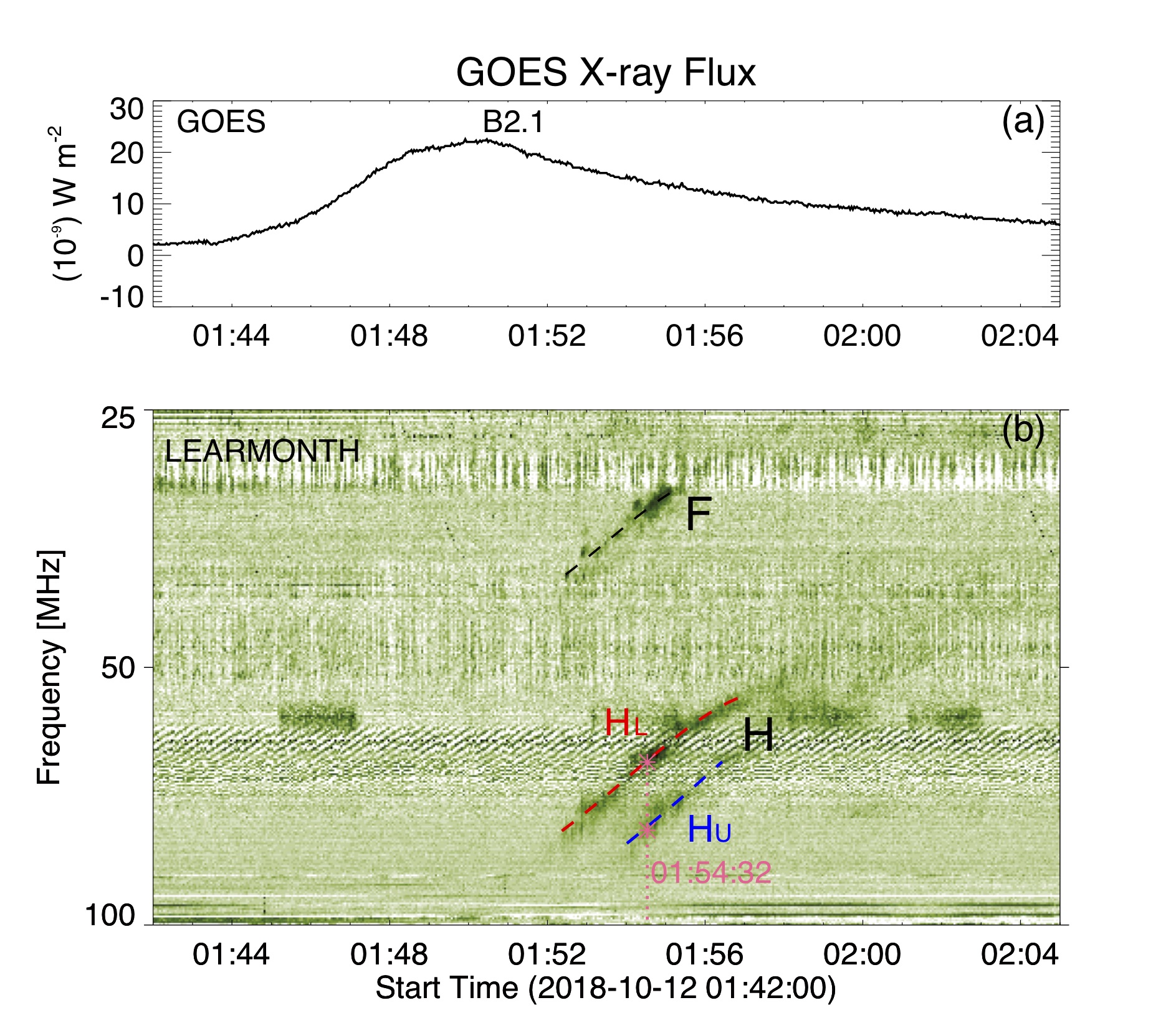}
\end{center}
\caption{ GOES soft X-ray flux during Flare1 in 1$\sim$8 \AA\ (a)  and a radio dynamic spectrum from Learmonth showing the information about the associated Type II radio burst (b). ``F'' indicates the fundamental frequency band and ``H'' mark the harmonic frequency band with the lower (``H$_{L}$'') and higher (``H$_{U}$'') splitting branches. The three curves along the fundamental (black), lower (red) and higher (blue) harmonic splitting bands are used to measure the frequency drifts and the height and speed of Shock1. 
The vertical line in panel (b) indicate the time at 01:54:32 UT.
The two pink asterisks mark the frequencies which are applied to calculate the density compression of Shock1.}\label{fig:2}
\end{figure}

\begin{figure}[ht!]
\begin{center}
\includegraphics[width=15cm]{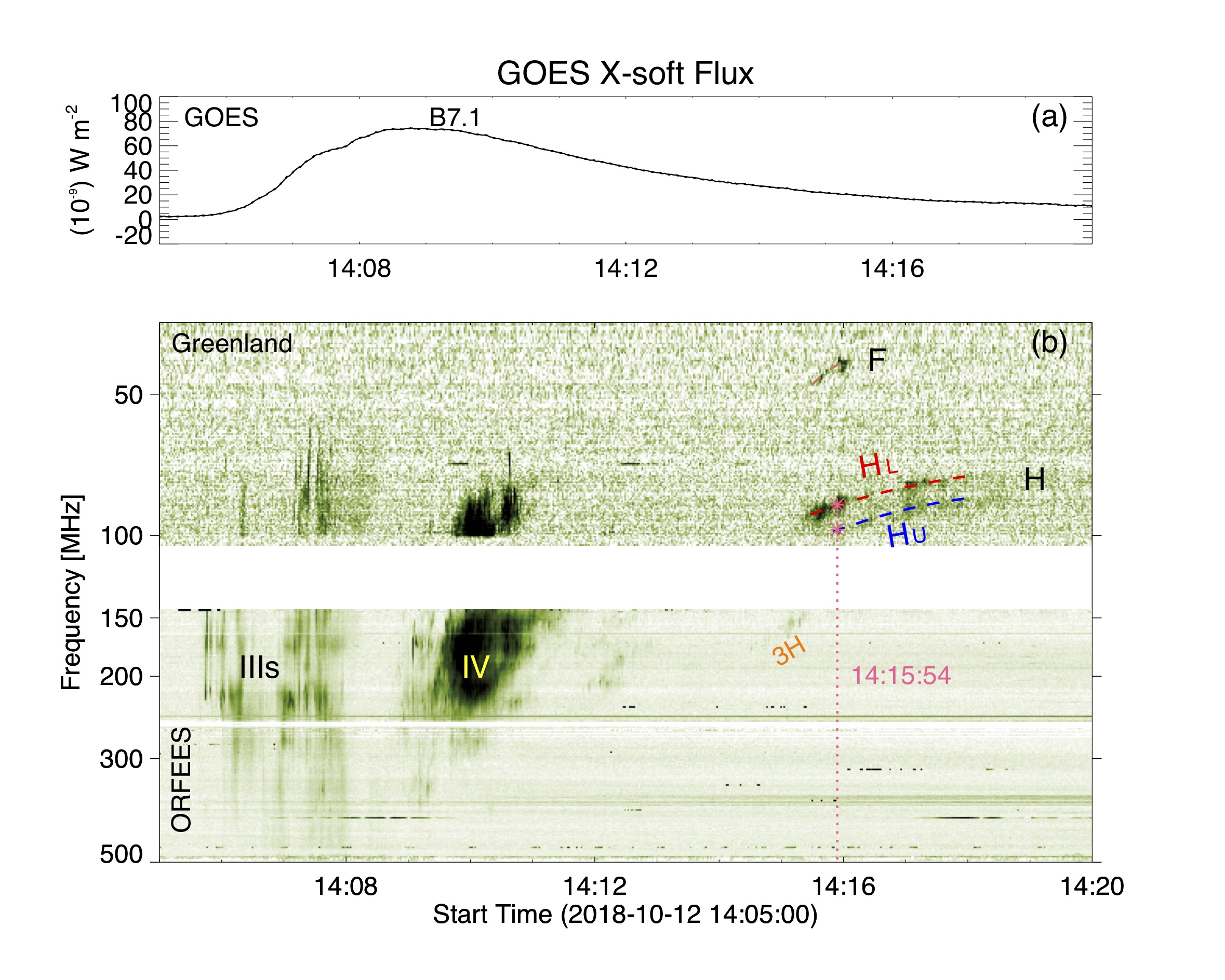}
\end{center}
\caption{GOES soft X-ray flux during Flare2 in 1$\sim$8 \AA\ (a) and the composite radio dynamic spectrum from Greenland (25$\sim$105 MHz) and ORFEES (144$\sim$400 MHz)  displaying the complicated radio bursts during E2 (b). 
Along the red and blue dashed curves in (b) we will measure the frequency drift of the Type II radio burst and derive the height and speed of the associated shock (Shock2). 
The vertical dashed line indicate the time at 14:15:54 UT.
The two pink asterisks mark the frequencies which are used to calculate the density compressions of Shock2.}\label{fig:3}
\end{figure}

The GOES X-ray flux of the B7.1-class flare (Flare2) and the composite dynamic spectrum from ORFEES (144$\sim$400 MHz) and Greenland (25$\sim$105 MHz) associated with E2 are displayed in the top and bottom panels of Figure~3, respectively.
A variety of radio bursts, such as a group of type III bursts (IIIs), a relatively strong short-lived type IV burst and a weak type II burst, can be found in the composite dynamic spectrum.
The Type III bursts occurred in the initial and impulsive phase of Flare2. 
Some of them show negative frequency drifts which probably result from energetic electron beams propagating outward along open coronal magnetic field \citep[e.g.,][]{Yan06,Huang11}, while others show positive drifts which may be caused by energetic electron beams propagating downward from where they are accelerated (likely the reconnection region) \citep[e.g.,][]{Reid14,Ning16,Tan16}.
The type IV burst appeared after the peak of Flare2 and lasted for about 2 minutes. 
It is probably excited by the energetic electrons trapped within the erupting magnetic structures \citep[e.g.,][]{Smerd71,Vlahos82,Stewart85}.

At around 14:15 UT on Oct 12, the type II burst appeared with obvious fundamental (F) and second harmonic bands (splitting into two lanes H$_{L}$ and H$_{U}$) in the observation of Greenland.
Similar to E1, it likely indicates a shock (Shock2) induced during the eruption E2.
Along H$_{L}$, the frequency varies from 90 MHz to 74 MHz with a mean frequency drift of -0.099 MHz s$^{-1}$ and that decreases from 98 MHz to 83 MHz with a mean frequency drift of -0.113 MHz s$^{-1}$ along H$_{U}$.
In addition, a very faint third harmonic band (3H) seems to appear in the dynamic spectrum of ORFEES. 
%
Using the same method described above (see Equation~(1)), we obtained the density jump of Shock2 at 14:15:54 UT (marked by the vertical line in Figure~3(b), when the H$_{U}$ is 98 MHz and H$_{L}$ is 86 MHz), which is about 1.30. The corresponding shock Mach number was deduced as 1.23.

\begin{figure}[ht!]
\begin{center}
\includegraphics[width=10cm]{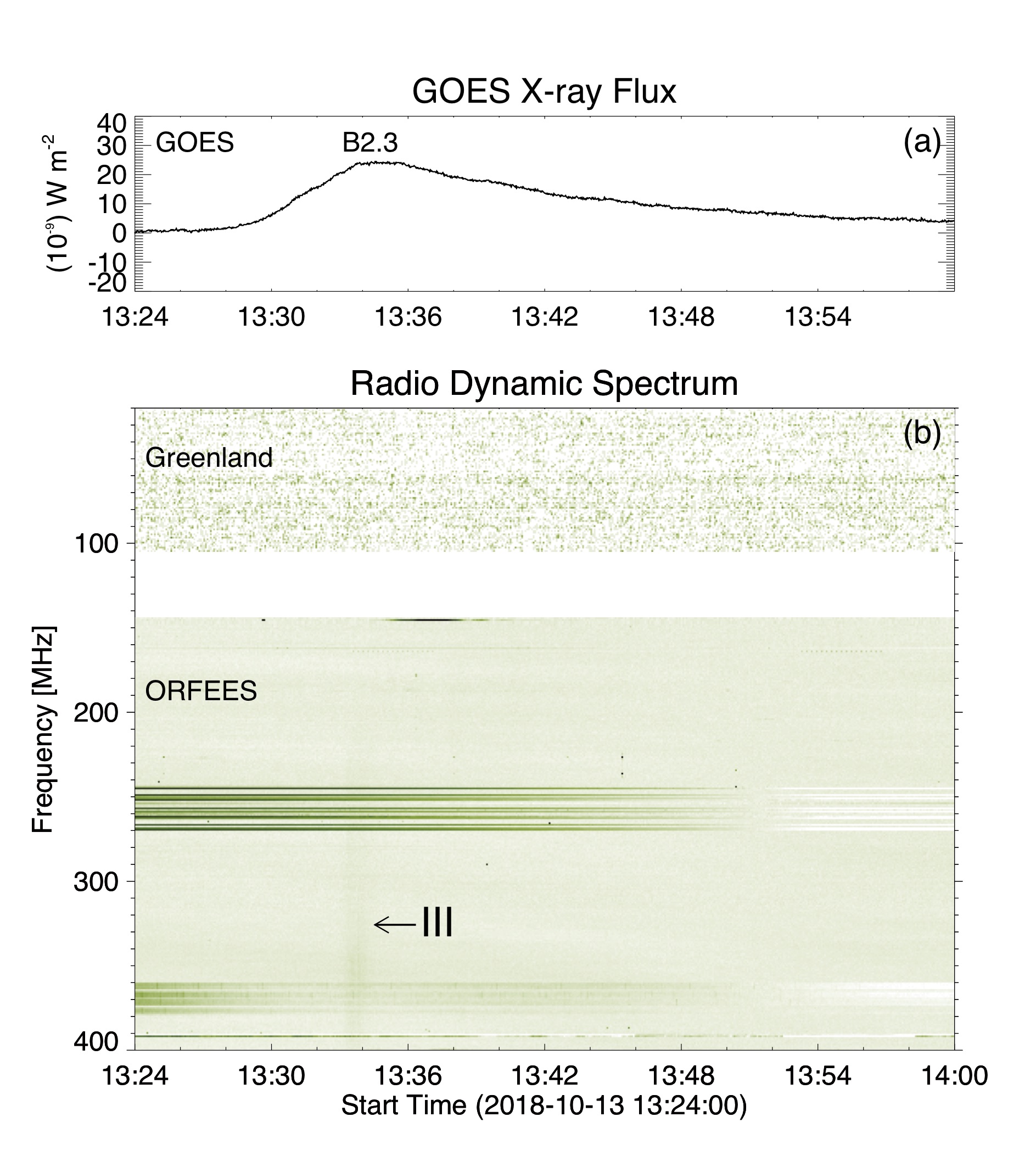}
\end{center}
\caption{GOES soft X-ray flux in 1$\sim$8 \AA\ during Flare3 (a) and a composite radio dynamic spectrum from Greenland (25$\sim$105 MHz) and ORFEES (144$\sim$400 MHz) during E3 (b). 
}\label{fig:4}
\end{figure}

\begin{figure}[ht!]
\begin{center}
\includegraphics[width=15cm]{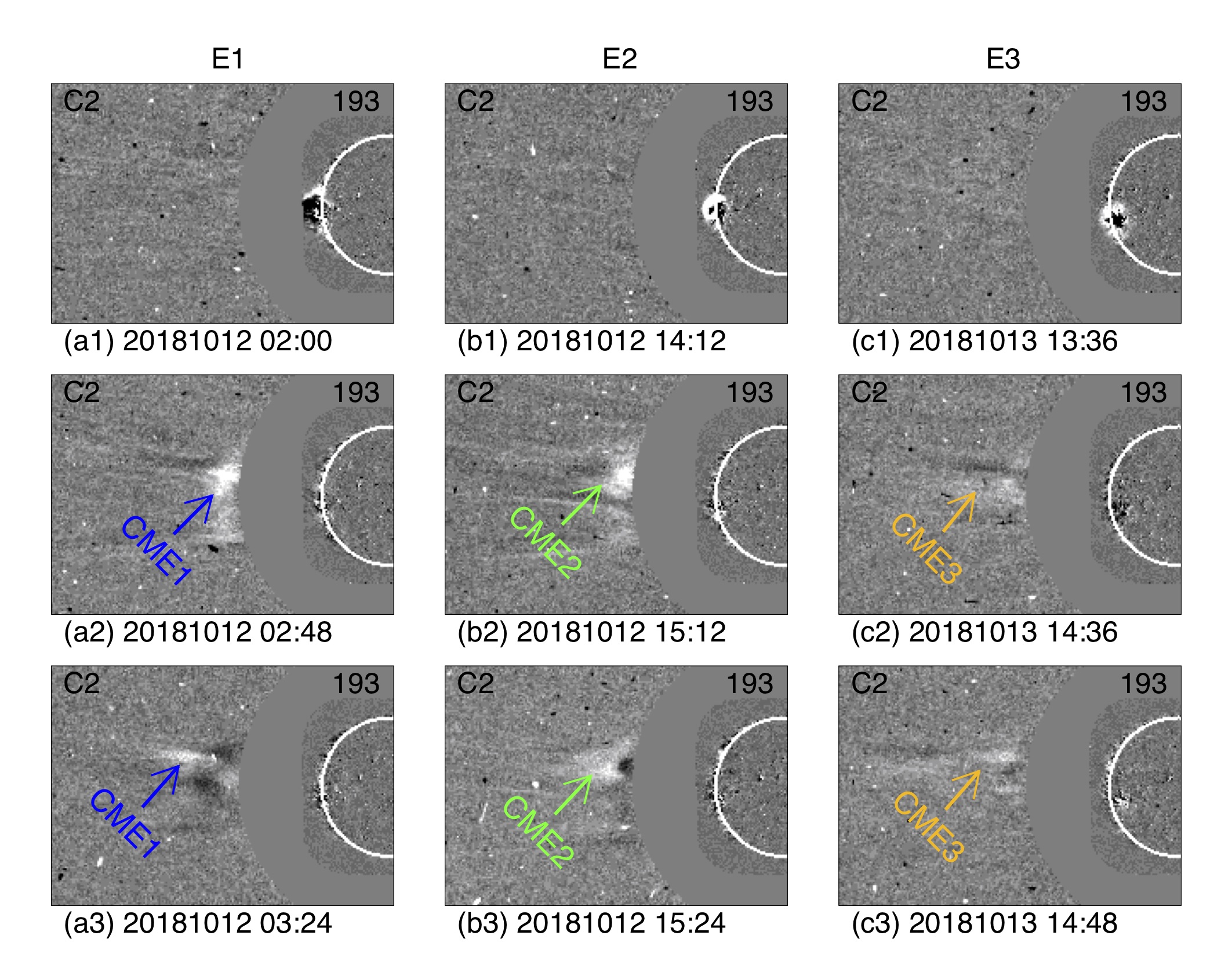}
\end{center}
\caption{The composite images of LASCO C2 WL and AIA 193 \AA\ running difference images show the propagations of the three faint CMEs. left column (a1-a3) for CME1, middle column (b1-b3) for CME2 and right column (c1-c3) for CME3.}\label{fig:5}
\end{figure}

Figure~4 shows the GOES X-ray flux of the flare (top panel) and the dynamic spectra (bottom panel) during E3. 
A B2.3-class flare (Flare3) correlated with E3.
Taking advantage of the joint observations from ORFEES and Greenland, we found that only a very weak type III burst and no sign of type II burst appeared in this event.\\

\subsection{White Light CMEs}
The running difference intensity images of LASCO C2 and AIA at 193 \AA\ are composited and shown in Figure~5, displaying the coronal changes during E1, E2, and E3.
Three faint stream-like CMEs (CME1--CME3) can be separately identified in the three eruptions.
They are indicated by the arrows in the left, middle and right column of Figure~5, respectively.
According to the LASCO CME catalog (see https://cdaw.gsfc.nasa.gov/CME$\_$list/index.html), CME1--CME3 only appear in the FOV of C2 (2$\sim$6R$_\odot$) and belong to poor events (quality index 1). 
They separately have an angular width of about 46$^\circ$, 44$^\circ$, and 38$^\circ$, and a 
2nd-order speed at the time of final height measurement of 333 km s$^{-1}$, 492 km s$^{-1}$, and 133 km s$^{-1}$(see Table~1).
The times of their first appearance in the LASCO C2's FOV are around 02:48 UT (Oct 12), 15:12 UT (Oct 12), and 14:36 UT (Oct 13), respectively.
Obviously, the Type II bursts in E1 and E2 had already formed before their corresponding CMEs came into the C2 FOV.
No relevant CME can be found from the observations of COR~1 (inner coronagraph) and COR~2 (outer coronagraph) on-board $STA$.
This may be related to the on-disk perspective of $STA$ and the weak magnitudes of the eruptions.
\\
\\
\subsection{Eruptions in EUV}

The AIA 193 \AA\ (Figure~6) and EUVI 195 \AA\ (Figure~7) intensity images display the early evolutions of E1--E3 from different perspectives.
The original images in the left columns of Figures~6,7 are utilized as references to get the base difference images in the middle and right columns.
The AIA 193 \AA\ data clearly exhibit the dome-like structures of CME1 and CME2 with distinct leading edges (LE1 and LE2) in their early stages, while CME3 had a tenuous leading edge (LE3).
Comparing the main eruption directions of E1--E3 (indicated by the yellow arrows in Figure~6) with the radial directions (denoted by the purple arrows in Figure~6), it can be found that CME1 and CME2 primarily propagated northeast, whereas CME3 was ejected approximately along the radial direction.

In the EUVI observations, three diffusing EUV waves (W1--W3) can be observed, as indicated by the orange arrows in Figure~7. 
EUV waves are also called ``EIT waves'' or global coronal waves, which are large-amplitude waves initially driven by the rapid lateral expansion of a CME in the low corona and later propagating freely \citep[cf.][]{Long17}. 
\citet{Ma09} showed the evidence that EUV wave front includes contribution from its associated CME at the early stage. More information about the ``EUV waves'' could be seen in the recent reviews \citep{Liu14,Warmuth15,Chenpf16,Long17}. 
The propagation directions of W1 and W2 are similar and mainly towards the north from their eruption centers, while the traveling of W3 has no obvious preference.
Combining the AIA and EUVI observations with a separation angle of 105 $^\circ$, we estimate that the eruption directions of E1 and E2 are alike along the $\sim$45$^\circ$ north of their radial directions and E3 basically erupted radially.

Interestingly, a special brightening area (``SBA'', Figures~6b2,~7b2) was observed at the north border of AR 12724 when the north flank of CME2 swept there (from $\sim$14:09 UT to $\sim$14:11 UT on Oct 12). It showed as an arc structure in the AIA images and a brow-like brightening in the EUVI images. The SBA may be caused by the interaction of CME or EUV waves with some coronal structures.
Later, another expanding dome-like structure and a propagating diffusing wave front can be detected to propagate forward through SBA in the AIA and EUVI intensity images, which are indicated by the blue arrows with ``2ND'' in Figures~6b3,b4,~7b3, respectively. 
It is probably a secondary wave.



\begin{figure}[ht!]
\begin{center}
\includegraphics[width=15cm]{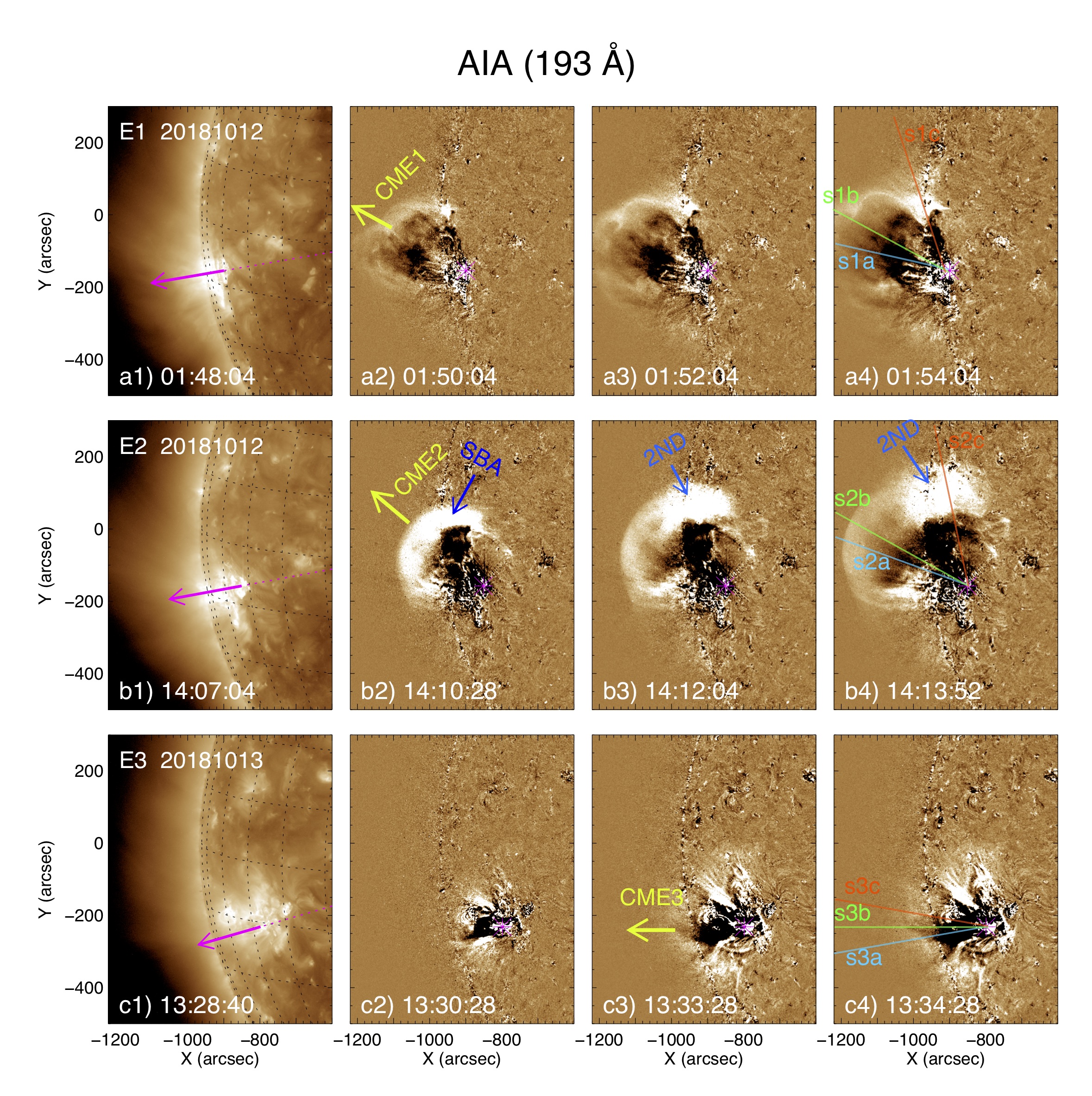}
\end{center}
\caption{The AIA 193 \AA\ intensity images display the early evolutions of E1 (top), E2 (middle), and E3 (bottom). 
The panels in left column (a1-a3) are original images and the rest panels (a2-a4, b2-b4,c2-c4) are base difference images. The pink arrows in (a1–a3) denote the heliocentric radial directions that passing through the eruption centers of E1, E2, and E3, respectively. The pink asterisks in panels (a2-a4, b2-b4,c2-c4) mark the centers of the eruptions. The yellow arrows in (a2,b2,c3) show the main eruption directions of CME1, CME2, and CME3, respectively. “SBA” (b2) and “2ND” (b3-b4) refer to a special brightening area and a secondary disturbance during E2. The lines in (a4,b4,c4) indicate the position of  slits s1a–s1c, s2a–s2c, and s3a–s3c, respectively. 
} \label{fig:6}
\end{figure}

\begin{figure}[ht!]
\begin{center}
\includegraphics[width=15cm]{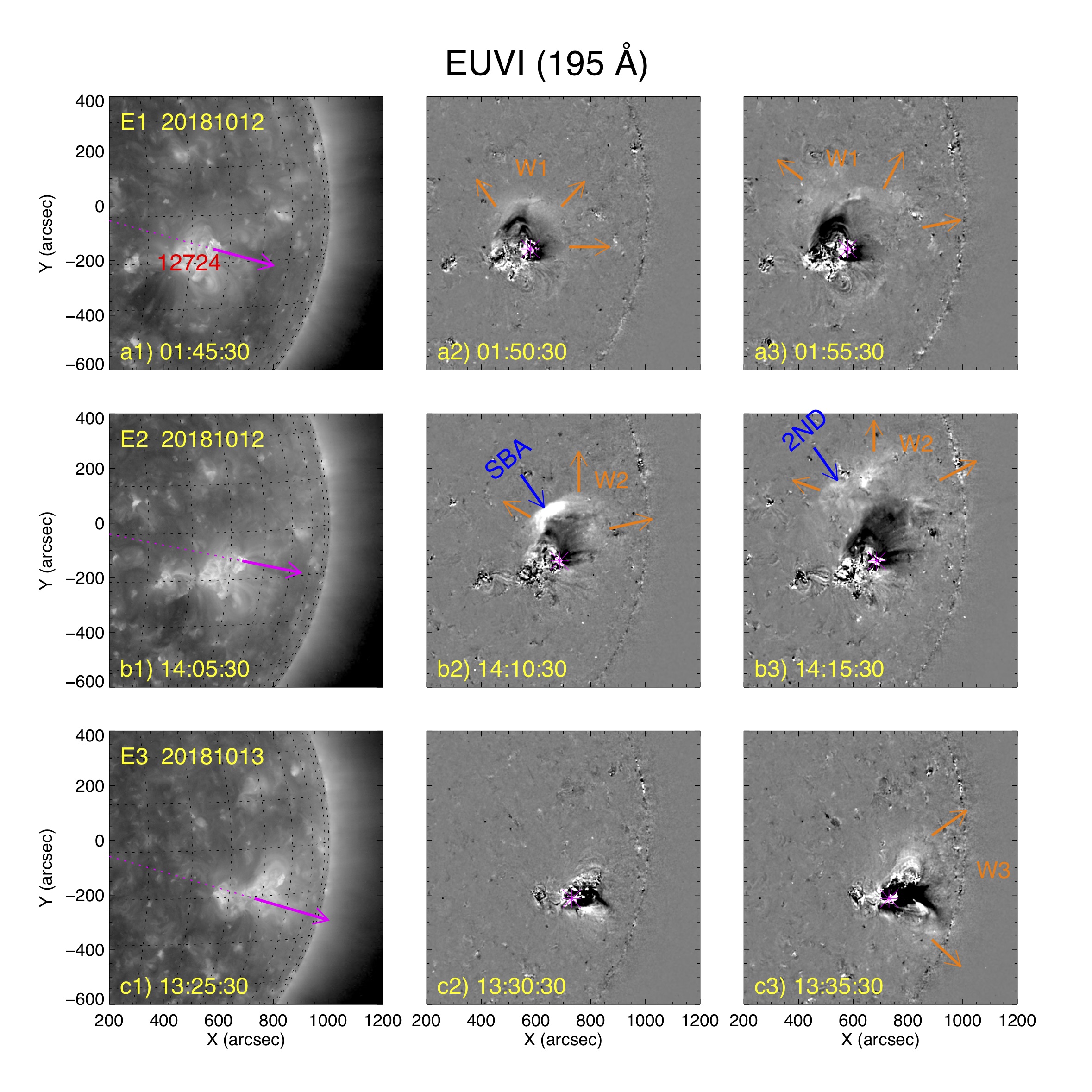}
\end{center}
\caption{The EUVI 195 \AA\ intensity images show the evolutions of E1 (a1-a3), E2 (b1-b3), and E3 (c1-c3) from on-disk perspective. 
The left column (a1,b1,c1) are the original images and the rest panels give the base difference images. Same to Figure~6, the pink arrows in the left column images show the heliocentric radial directions through each eruption centers (pink asterisks)
The orange arrows indicate the propagation directions of  the associated EUV wave ``W1'' (a2,a3)  ``W2'' (b2,b3) and ``W3'' (c3).} \label{fig:7}

\end{figure}
\begin{figure}[ht!]
\begin{center}
\includegraphics[width=15cm]{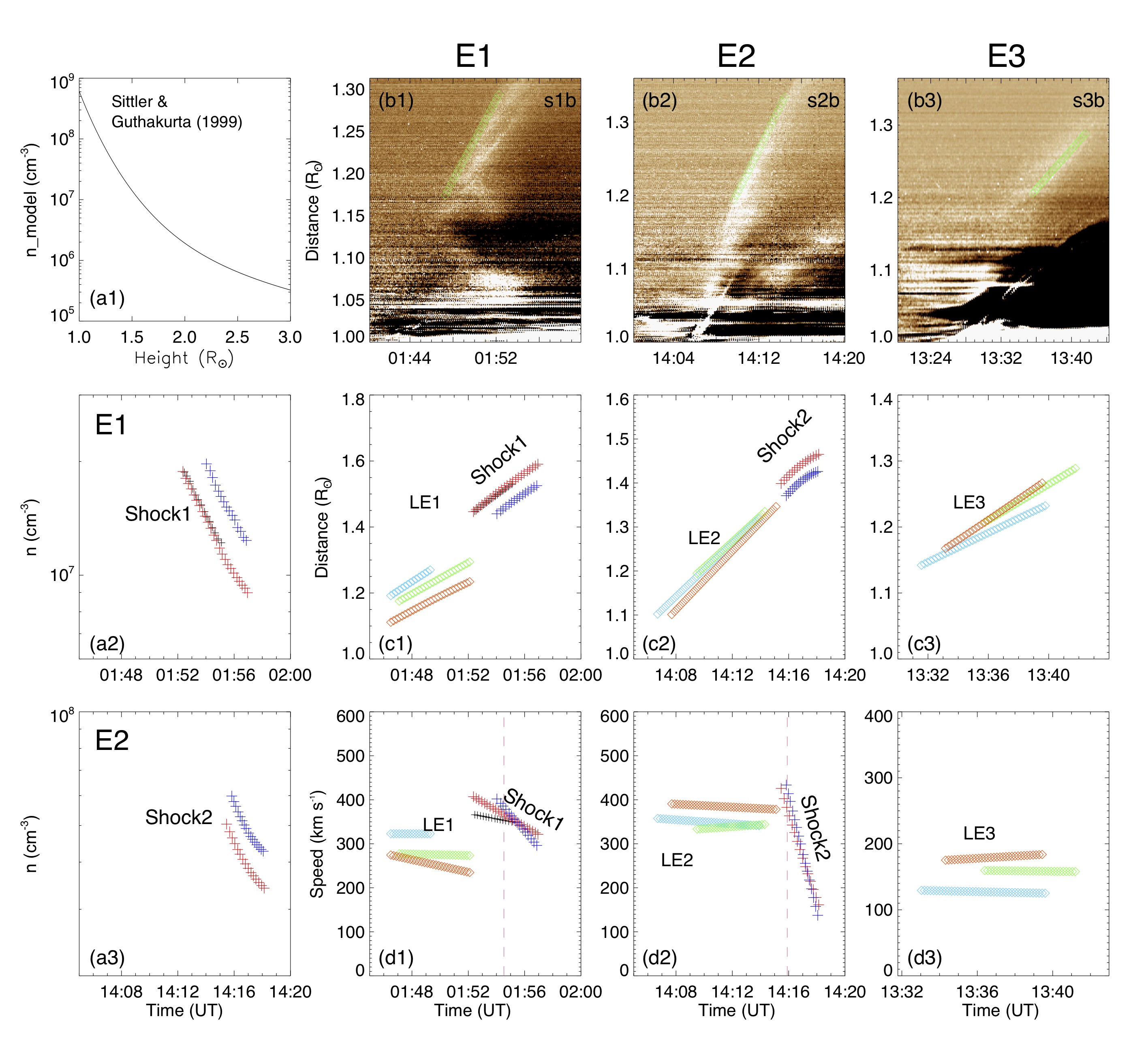}
\end{center}
\caption{
(a1) Shows the electron density changing with the height base on the model of \citet{Sittler99}. (a2,a3) Display the local plasma electron densities of Shock1 and Shock2 calculated from the observations of their associated Type II radio bursts. 
(b1--b3) Are the time-slit images showing the motions of LE1, LE2 and LE3 along the slits s1b, s2b and s3b, respectively. 
(c1--c3) Display the heliocentric distances of LE1 (diamonds) and Shock1 (pluses), LE2 (diamonds) and Shock2 (pluses), and LE3 (diamonds), respectively.
(d1--d3) Are the same to (c1--c3), but for the speeds of LE1--LE3 and Shock1--Shock2.
The identical colors in this figure represent the same bands or slits in Figure~2,~3,~6.
} \label{fig:8}
\end{figure}


\subsection{Kinetics of the CMEs and Shocks}
In order to explore the relationship between the CMEs and their associated Type II radio bursts, we studied the kinetics of CME1--CME3 and Shock1--Shock2, which are presented in Figure~8.
For convenience, the identical colors represent the same bands or slits in Figures~2,~3,~6,~8. 
Because there is no associated radioheliograph observation to be available, the exact locations of Type II burst sources are hard to be determined. 
Assuming that the electron density of the corona varies with heliocentric distance and the shock propagates along the radial direction, we first deduced the local plasma densities ($n$) of Shock1 and Shock2 from the observed frequency ($f_{p}$ indicated by the black dashed line in Figure~2 or 2$f_{p}$ indicated by the red and blue dashed lines in Figures~2 and 3) in the light of their relationship \begin{equation}
 f_{p}=8.98\times10^{3}\sqrt{n}
\end{equation}
The results are shown in Figures~8a2, a3. 
In order to further calculate the heights of Shock1 and Shock2 from the derived local plasma densities $n$, we apply the coronal plasma density model of \citet{Sittler99} (see also Figure~8a1), 
\begin{equation} n(z)=n_0\,a_1\,z^{2}\,e^{a_2\,z}[1+a_{3}\,z+a_{4}\,z^{2}+a_{5}\,z^{3}]
\end{equation}
\begin{displaymath}
z=1/(1+y), a_1=0.001292, a_2=4.8039,
\end{displaymath}
\begin{displaymath}
a_3=0.29696, a_4=-7.1743,a_5=12.321,
\end{displaymath}
where y is the height above the solar surface in solar radii and $n_0$ is the electron number density at the solar surface.
We choose $n_0$ as 6.0$\times$$10^8 $ cm$^{-3}$ considering the events under this study occurring in the period of solar activity minimum, which is similar to the value used in \citet{Ma11}.
The time distance profiles of Shock1 and Shock2 calculated from the different splitting bands in the radio dynamic spectra are presented in Figures~8c1,~c2, respectively.
It can be seen that the heights of Shock1 and Shock2 approximately change from 1.45 R$_\odot$ to 1.60 R$_\odot$ and from 1.37 R$_\odot$ to 1.46 R$_\odot$, respectively.

The leading edges (LE1--LE3) of CME1--CME3 can be tracked in the AIA running difference images at 193 \AA. 
For each event, we chose three different slits (s1a--s1c, s2a--s2c, and s3a--s3c in Figure~6) to make the time-distance slit images, which display the propagations of the leading edges along different directions.
As examples, three time-distance diagrams from s1b, s2b and s3b are plotted in the panels (b1--b3) of Figure~8.
According to the tracks or stripes in the slit images, we calculated the heights of LE1--LE3 and derived their speeds, which are shown by the diamonds in Figures~8c1--c3,d1--d3, respectively.

From Figure~8c1, it can be seen that although LE1 had moved out of the AIA's FOV when Shock1 began to appear, the development trend of the LE1's heights and the height variation of Shock1 suggest a high degree correlation between them. 
Figure~8c2 indicates a similar situation for LE2 and Shock2.
Thus, it is likely that the two shocks corresponding to the two Type II radio bursts in E1 and E2 were separately triggered by the expanding of the leading edges of CME1 and CME2.
In Figures~8d1,d2, it can be found that LE1 and LE2 have various speeds along different propagation directions. LE1 has the largest velocity of $\sim$320 km s$^{-1}$ (azure diamonds in Figure~8d1) along s1a (azure line in Figure~6a4). 
The fastest speed of LE2 is $\sim$380 km s$^{-1}$ (red diamonds in Figure~8d2), which was calculated along the slit s2c (red line in Figure~6b4). 

The speeds of Shock1 and Shock2 derived from their heights are given in Figures~8d1,d2, respectively.
It can be found that the shocks' speeds calculated along the different Type II bursts bands are also different.
These different speeds represent the speeds of the downstream (blue pluses) and upstream (red and black pluses) shock regions, which might be distinct from each other.
In addition, the discrepancies of the speeds are also probably caused by the measurement errors.
On average, Shock1 has an initial speed of $\sim$400 km s$^{-1}$ and that of Shock2 is $\sim$430 km s$^{-1}$.
According to the relationship between the shock's speed ($V_{s}$) and the local \Alfven\ speed $V_{A}$, i.e.,
\begin{equation}
M_{A}=\frac{V_{s}}{V_{A}}
\end{equation}
the \Alfven\ speeds at the early phases of Shock1 (01:54:32 UT, indicated by the vertical line in Figures~2b,~8d1) and Shock2 (14:15:54 UT, indicated by the vertical line in Figures~3b,~8d2) can be deduced as $\sim$300 km s$^{-1}$ and $\sim$350 km s$^{-1}$, respectively.
Compared with the fastest speeds of LE1 ($\sim$320 km s$^{-1}$) and LE2 ($\sim$380 km s$^{-1}$), the local \Alfven\ speeds are smaller.
These results are in agreement with the scenario of piston-driven shock, supporting our conjecture that the CMEs in E1 and E2 excited their relevant shocks and Type II radio bursts.
Figure~8d3 shows that the leading edge of CME3 had a relatively slower speed ($\sim$160 km s$^{-1}$) than LE1 and LE2.
It is likely less than the local \Alfven\ speed, which might be the reason why Type II radio burst or shock is absent in E3.
\\
\subsection{Background Fields from PFSS Extrapolation}
To probe the background field structures surrounding the eruption source area and their relationship with the eruptions, an extrapolation was performed using the potential field source surface (PFSS) model \citep[e.g.,][]{Schatten69,Schrijver03} with a starting radius of 1.01 \Rsun. 
For better reliability, one HMI longitudinal magnetogram on 2018 Oct 13 was applied
to extrapolate the potential field.
The magnetic field lines from the extrapolation are overlaid on the AIA 171 \AA, EUVI 195 \AA\ intensity images, HMI and rotated HMI magnetograms, as displayed by Figures~9a--d.


The HMI magnetogram in Figures~9d shows that AR 12724 mainly consists of the leading positive flux ``P'' and following negative flux ``N'', with some surrounding parasitic magnetic elements, such as the fluxes ``n'' and ``p''.
In this panel, we also overlaid the profiles of the three filaments F1--F3, which are indicated by the red curves.
It can be seen that F1 and F2 are located at the northwest of the AR and aligned along the magnetic neutral lines between N and p, while F3 lies in the AR's southeast region between the opposite polarity fluxes P and n.
The spine directions of F1 and F2 are approximately from north to south, opposite to the east-west orientation of F3.

At the remote region to the north of AR 12724, we found some different magnetic loop systems, which are represented by the short dark blue field lines.
Between these magnetic systems and the magnetic loops inside AR 12724, the magnetic quasi-separatrix layer (indicated by the yellow dashed line) may exist.
That area is consistent with the place where the special brightening region SBA in E2 appeared.
It is likely that SBA was caused by the interaction between the north flank of CME2 and the magnetic separatrix layer.



\begin{figure}[ht!]
\begin{center}
\includegraphics[width=15cm]{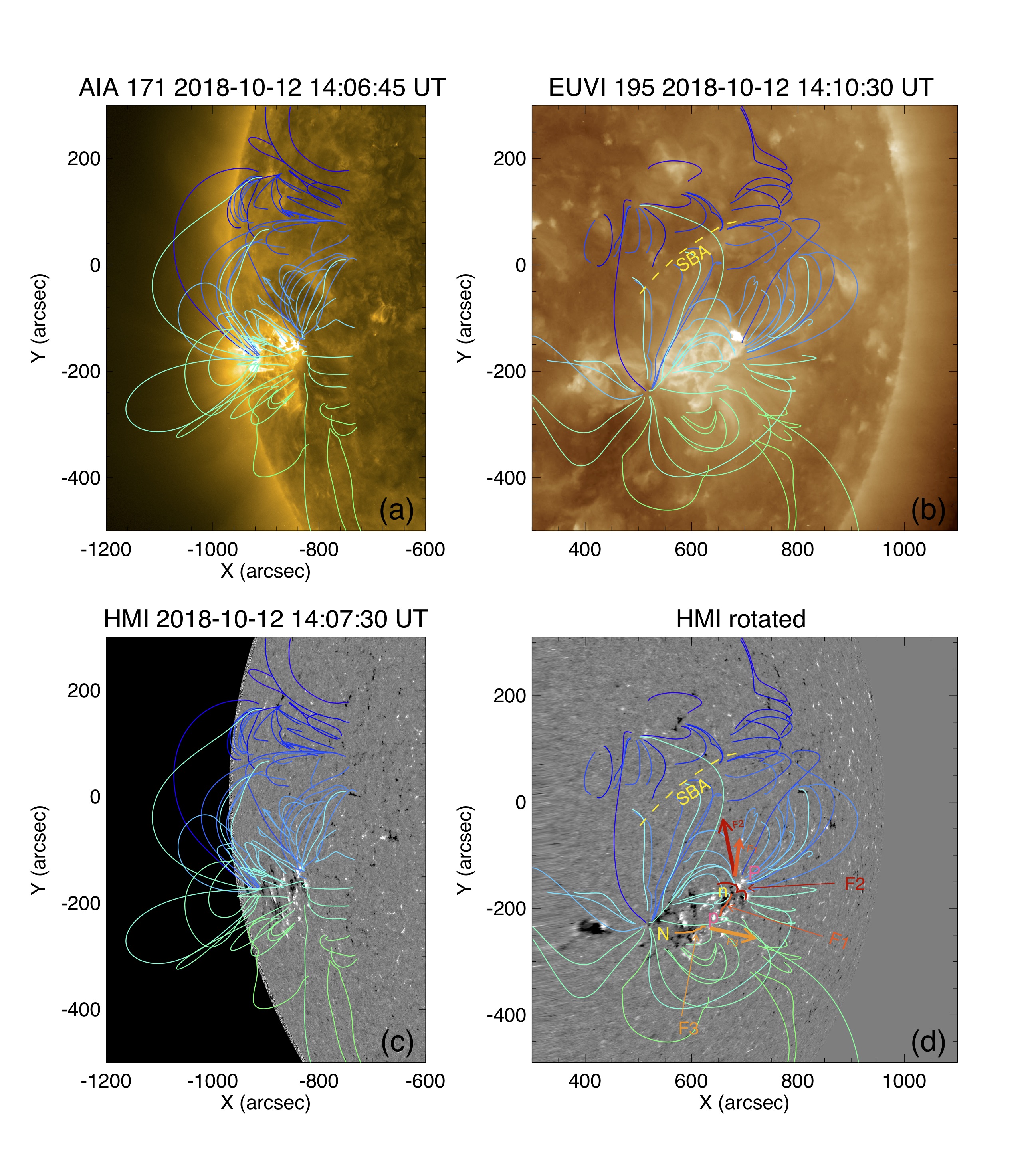}
\end{center}
\caption{The extrapolated magnetic field lines using the PFSS model are overlaid on the AIA, EUVI intensity images and HMI magnetograms. The yellow dashed curves in the panels (b) and (d) indicate the location of the ``SBA''. 
``P'' and ``N'' mark the leading and following magnetic fluxes of AR 12724.
 ``p'' and ``n'' denote the positive and negative parasitic magnetic elements surrounding N and P, respectively.
The thick arrows point to the main eruption directions of F1--F3.
}\label{fig:9}
\end{figure}

\section{Summary and Discussion}

Using radio dynamic spectra and dual perspective EUV observations, We investigate three successive solar eruptions (E1, E2 and E3) from the same active region AR 12724. 
All the eruptions were accompanied by a B-class flare and a slow faint WL CME. 
However, only the first two of them were observed to correlate with Type II radio bursts, suggesting the likely appearances of shocks (Shock1 and Shock2) only in the two events. 
From the radio dynamic spectra, we utilize the splitting bands of the Type II bursts to estimate the density jumps (1.44 and 1.30),  \Alfven\ Mach numbers (1.35 and 1.23), and coronal \Alfven\ speeds ($\sim$300 km s$^{-1}$ and $\sim$350 km s$^{-1}$) of Shock1 and Shock2.
Through a comparative study, we found that 
\begin{itemize}
\item The apparent speeds of the CMEs' leading edges (LE1--LE3) are different. LE3 has an obvious slower speed ($\sim$160 km s$^{-1}$) than LE1 ($\sim$320 km s$^{-1}$) and LE2 ($\sim$380 km s$^{-1}$).
The speeds of LE1 and LE2 can exceed their corresponding local \Alfven\ speeds ($\sim$300 km s$^{-1}$ and $\sim$350 km s$^{-1}$).
\item E1 and E2 originated from the northwest of AR 12724, while E3 took place from the AR's southeast region.  
\item The EUV imaging observations from two different perspective indicate that E1 and E2 erupted along the $\sim$45$^\circ$ north of their radial directions, while CME3 in E3 approximately propagated radially.
\end{itemize}

\subsection{Trigger of Type II Radio Bursts}

According to the $GOES$ soft X-ray flux data, Flare3 is a B2.3-class flare and stronger than the B2.1 Flare1. 
However, Type II radio burst is associated with the weaker one.
This is in agreement with the finding that the magnitudes of flares are not directly related to the occurrence of Type II radio bursts \citep[e.g.,][]{Cliver99}. In addition, since all the eruptions took place from the same AR and all the associated flares are relatively weak (only B-class), it is hard to conclude that the Type II bursts studied here were initiated by the blast wave due to flares.
Our calculations have shown that the speeds of LE1 and LE2 along certain directions can exceed the local coronal \Alfven\ speeds, which meet the requirements of the formation of a piston-driven shock.
Thus, It would be more reasonable that the Type II bursts were triggered by their associated CMEs.
On the other, it should be noted that the third eruption E3 has a different source region and eruption direction from E1 and E2.
The coronal plasma and magnetic field environments that the erupting structures of E3 encountered would be also distinct from those of E1 and E2. The missing of Type II burst in E3 may be associated with this situation as well.\\

\subsection{SBA and the Secondary Wave}
In the eruption E2, a special brightening area SBA is detected where a magnetic separatrix may exist according to the results of the PFSS extrapolation (Figure~9).
The occurrence or appearance of SBA in this event might be explained by this scenario:
when the flank of CME2 and/or EUV wave W2 arrived the magnetic separatrix layer, it would be likely compressed and heated, which might give birth to SBA.  
Along with the occurrence of the reflection and refraction of the EUV wave near the magnetic separatrix, a secondary wave 2ND might be further produced and propagate outward. 
Similar situations can be found in some other studies \citep[e.g.,][]{Ofman02,Shen12,Chandra16,Zheng18}.
In addition, some studies have shown a close relationship between Type II radio bursts and  such interactions \citep[e.g.,][]{Kong12,Feng12,Shen19}.
Using simultaneous radio and EUV imaging data, \citet{Cheny14} found that the source location of a solar type II radio burst coincides with the interface between CME EUV wave front and a nearby coronal ray structure, where an obvious EUV brightening also appeared.
They conjectured that the CME–streamer interactions may be important to the formation of type II radio burst.
Unfortunately, there is no radioheliograph observation available for our study and we can not confirm the exact location of the type II burst in E2, but according to the results of \citet{Cheny14}, it can be suspected that the special brightening area SBA corresponds to the source region of the type II burst in E2.
\section*{Acknowledgments}
We thank Prof. Jun Lin and Prof. Baolin Tan for their insightful suggestions and informative discussions. We are grateful to the referees for their constructive comments and suggestions. We acknowledge NASA’s open data policy in using SDO data. STEREO is a mission in NASA’s Solar Terrestrial Probes program. SOHO is a project of international collaboration between ESA and NASA. We are grateful to ORFEES, Learmonth and Greenland teams for offering the radio spectra data. This work was supported by NSFC (11433006, 11533008,11661161015,11790300, 11790301, 11790304, 11941003, 11973057 and 41331068) and the B-type Strategic Priority Program of the Chinese Academy of Sciences, Grant No. XDB41000000.

\bibliographystyle{aasjournal}
\bibliography{ms_20181012_63}



\end{CJK*}
\end{document}